\documentclass[twocolumn,showpacs,preprintnumbers,superscriptaddress,amsmath,%
amssymb,nofootinbib]{revtex4}

\input psfig.sty
\usepackage{epsfig}

\usepackage{graphicx}
\usepackage{dcolumn}
\usepackage{bm}

\newcommand{\be}{\begin{equation}}
\newcommand{\ee}{\end{equation}}
\begin{document}

\title{New coupled quintessence cosmology}

\author{J. F. Jesus}\email{jfernando@astro.iag.usp.br}

\author{R. C. Santos}\email{cliviars@astro.iag.usp.br}

\affiliation{Universidade de S\~{a}o Paulo, 05508-900, S\~ao Paulo,
SP, Brasil}

\author{J. S. Alcaniz}\email{alcaniz@on.br}

\affiliation{Observat\'orio Nacional, 20921-400, Rio de Janeiro --
RJ, Brasil}

\affiliation{Instituto Nacional de Pesquisas Espaciais/CRN,
59076-740, Natal -- RN, Brasil}

\author{J. A. S. Lima}\email{limajas@astro.iag.usp.br}

\affiliation{Universidade de S\~{a}o Paulo, 05508-900, S\~ao Paulo,
SP, Brasil}

\vspace{0.1 cm}

\date{\today}

\begin{abstract}
A component of dark energy has been recently proposed to explain the
current acceleration of the Universe. Unless some unknown symmetry
in Nature prevents or suppresses it, such a field may interact with
the pressureless component of dark matter, giving rise to the
so-called models of coupled quintessence. In this paper we propose a
new cosmological scenario where radiation and baryons are conserved,
while the dark energy component is decaying into cold dark matter
(CDM). The dilution of CDM particles, attenuated with respect to the
usual $a^{-3}$ scaling due to the interacting process, is
characterized by a positive parameter $\epsilon$, whereas the dark
energy satisfies the equation of state $p_x=\omega \rho_x$ ($\omega
< 0$). We carry out a joint statistical analysis involving 
recent observations from type Ia supernovae, baryon acoustic
oscillation peak, and Cosmic Microwave Background shift parameter to
check the observational viability of the coupled quintessence
scenario here proposed.

\end{abstract}


\maketitle

\section{Introduction}

According to Einstein's general theory of relativity, the dynamic
properties of a given spacetime are determined by its total energy
content. In the cosmological context, for instance, this amounts to
saying that to understand the spacetime structure of the Universe
one needs to identify the relevant sources of energy and their
contributions to the total energy momentum tensor. Matter fields
(e.g., baryonic matter and radiation), are obvious sources of
energy. Nevertheless, according to current observations, two other
components, namely, dark matter and dark energy, whose origin and
nature are completely unknown, are governing the late time dynamic
properties of the Universe. Although fundamental to our
understanding of the Universe, several important questions involving
these dark components and their roles in the dynamics of the
Universe remain unanswered (see, e.g., \cite{review} for some recent
reviews).

Among these questions, the possibility of interaction in the dark
sector (dark matter-dark energy), which gave origin to the so-called
models of coupled quintessence, has been largely explored in the
literature \cite{cq,cq2}. These scenarios are based on the premise
that, unless some special and unknown symmetry in Nature prevents or
suppresses a non-minimal coupling between these components (which
has not been found -- see, e.g., \cite{carroll} for a discussion),
such interaction is in principle possible and, although no
observational piece of evidence has so far been unambiguously
presented, a weak coupling still below detection cannot be
completely excluded.

>From the observational viewpoint, these models are capable of
explaining the current cosmic acceleration, as well as other recent
observational results \cite{cq}. From the theoretical point of view,
however, critiques to these scenarios do exist and are mainly
related to the fact that in order to establish a model and study
their observational and theoretical predictions, one needs first to
specify a phenomenological coupling between the cosmic components.

In this concern, an interesting step towards a realistic interaction
law was given recently by Wang \& Meng in Ref. \cite{wm} (see also
\cite{alclim05}) in the context of models with vacuum decay,  a class
of coupled quintessence in which the dark energy equation of state
(EoS) is $w = -1$. Actually, in certain sense, one may say that
coupled dark energy or quintessence  models are the natural
inheritors of the so-called time-varying $\Lambda(t)$-cosmologies
\cite{Brons1,Brons2,list,list2}. However, instead of the traditional approach,
Refs. \cite{wm,alclim05} deduced a new interaction law from a simple
argument about the effect of the dark energy on the cold dark matter
(CDM) expansion rate. The resulting expression is a very general law
that has many of the previous phenomenological approaches as a
particular case.

\begin{figure*}
\centerline{\epsfig{figure=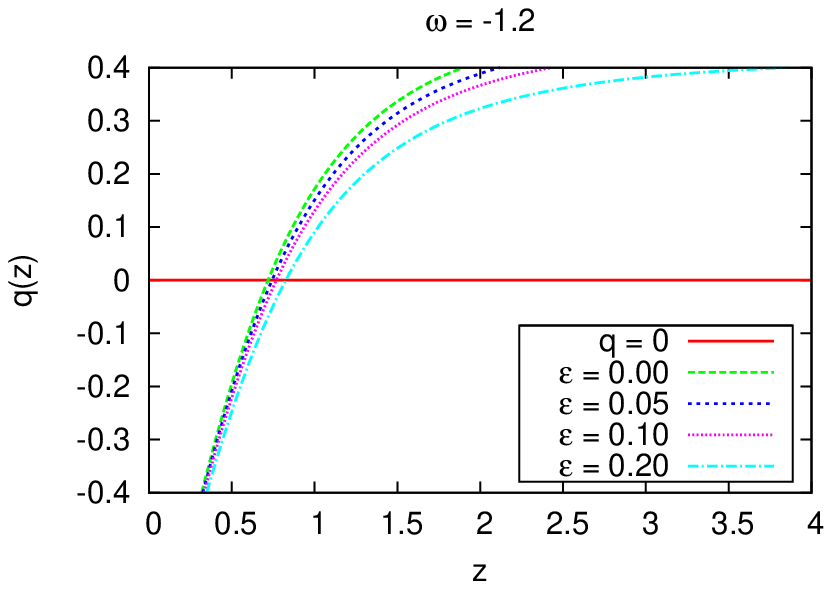,width=2.4truein,height=2.8truein}
\epsfig{figure=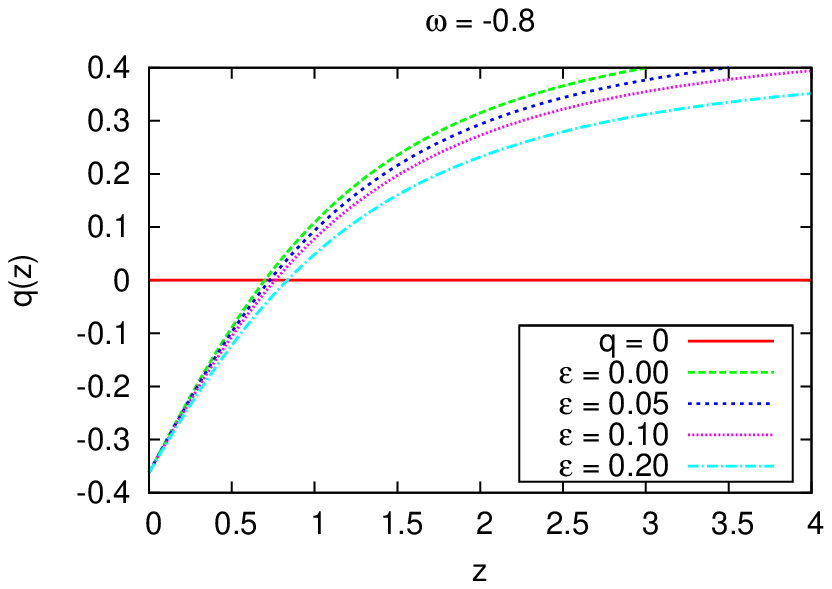,width=2.4truein,height=2.8truein}
\epsfig{figure=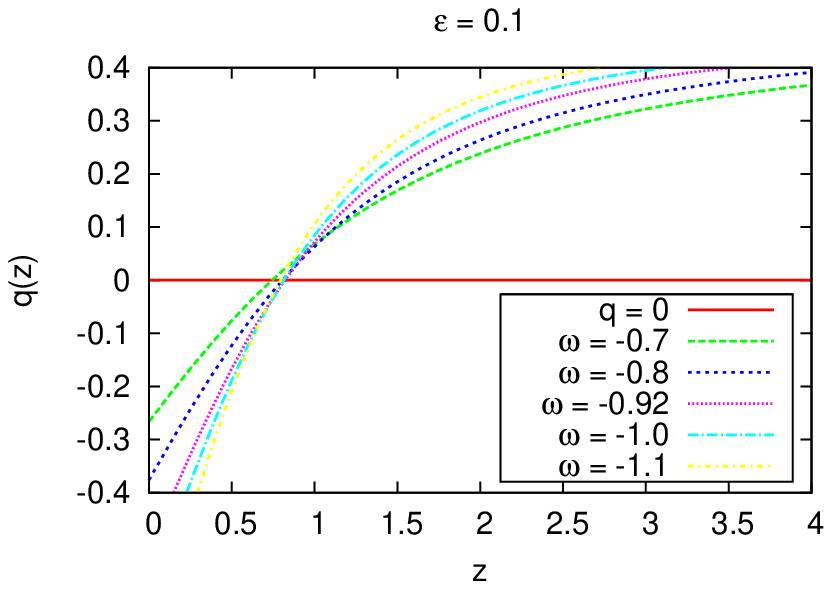,width=2.4truein,height=2.8truein} \hskip
0.1in} \caption{$q(z)$ in the scenario of coupled quintessence.
{\bf{a)}} Deceleration parameter as a function of redshift for the
phantom case in which $\omega=-1.2$ and selected values of
$\epsilon$. {\bf{b)}} The same as in the previous Panel for the
quintessence case $\omega=-0.8$. As discussed in the text, the
effect of a positive $\epsilon$ parameter is to decrease the value
of $q(z)$, therefore increasing the value of the transition redshift
$z_t$. $q(z)$ versus redshift for the specific value of
$\epsilon=0.1$ and selected values of EoS parameter.}
\label{fig:qzw}
\end{figure*}

In this paper, we extend the arguments of Refs. \cite{wm,alclim05}
to a dark energy/dark matter interaction, where the dark energy
component is described by an equation of state $p_x = \omega \rho_x$
($w < 0$), and explore theoretical and observational consequences of
a new scenario of coupled quintessence. Differently from other
interacting quintessence models, we do not consider interaction
between the dark sector and the baryonic content of the Universe. We
also emphasize that this process of interaction is completely
different from the physical point of view from unification scenarios
of the dark sector, an idea that has been widely discussed in the
recent literature \cite{chap}.

We have organized this paper as follows. In Sec. II the interaction
law and the basic field equations of the model are presented. The
influence of the dark energy-dark matter coupling on the epoch of
cosmic acceleration is also discussed. In order to test the
observational viability of the model, Sec. III presents a
statistical analysis involving the most recent type Ia supernovae
(SNe Ia) data \cite{davis,wv07,rnew,Astier06,Ries07}, observations
of the baryon acoustic oscillation (BAO) peak (measured from the
correlation function of luminous red galaxies) \cite{bao} and the
current estimate of the Cosmic Microwave Background (CMB) shift
parameter from WMAP-5 \cite{Sperg07}. In Sec. IV we end this paper
by summarizing our main results.

\section{The model}

For a spatially flat, homogeneous and isotropic scenario driven by
matter (baryonic + dark) and radiation fields and a
negative-pressure dark energy component, the Einstein field
equations can be written as
\begin{equation}
8\pi G(\rho_{\gamma} + \rho_b + \rho_{dm} + \rho_x) = 3\frac{{\dot
a}^2}{a^2}, \label{fried}
\end{equation}
\begin{equation}
\label{p} 8\pi G(p_{\gamma}+ p_x) =-2\frac{{\ddot a}}{a}
-\frac{{\dot a}^2}{a^2}\ ,
\end{equation}
where $\rho_{\gamma}$, $\rho_b$, $\rho_{dm}$ and $\rho_x$, are the
energy densities of the radiation, baryons, cold dark matter and
dark energy, respectively, while $p_{\gamma}$ and $p_x$ are  the
radiation and dark energy pressures. We will always refer to the
dark matter quantities with the subscript $(dm)$, in order to
distinguish them to total matter quantities, for which we will use
the subscript $(m)$. Thus, in our notation,
$\rho_m=\rho_b+\rho_{dm}$.

By assuming that the radiation and baryonic fluids are separately
conserved, the energy conservation law for the two interacting
components ($u_{\alpha}{\bar{T}^{\alpha\beta}}_{{;}\beta}=0$, where
$\bar{T}^{\alpha\beta} = T^{\alpha\beta}_{dm} + T^{\alpha\beta}_x$)
reads
\begin{equation}\label{coupling}
\dot{\rho}_{dm} + 3 \frac{\dot{a}}{a}\rho_{dm} = -\dot{\rho}_x -
3\frac{\dot{a}}{a}(\rho_x + p_x)\; .
\end{equation}

Now, to complete the description of our interacting quintessence
scenario we need to specify the interaction law. In principle, if
the quintessence component is decaying into CDM particles, the CDM
component will dilute more slowly compared to its standard
(conserved) evolution, $\rho_m \propto a^{-3}$. Therefore, if the
deviation from the standard evolution is characterized by a positive
constant $\epsilon$ we may write\footnote{The positiveness of the
interacting parameter $\epsilon$ is required from thermodynamical
arguments. For a discussion, see \cite{alclim05}.}
\begin{equation}
\rho_{dm}=\rho_{dm0}{a}^{-3 + \epsilon},
\end{equation}
where $\epsilon$ is a constant parameter and we have set the
present-day value of the cosmological scale factor $a_0 = 1$. In
what follows we also consider that the dark energy component is
described by an equation of state $p_x = \omega \rho_x$, where the
constant $\omega$ is a negative quantity.

\begin{figure*}
\centerline{\epsfig{figure=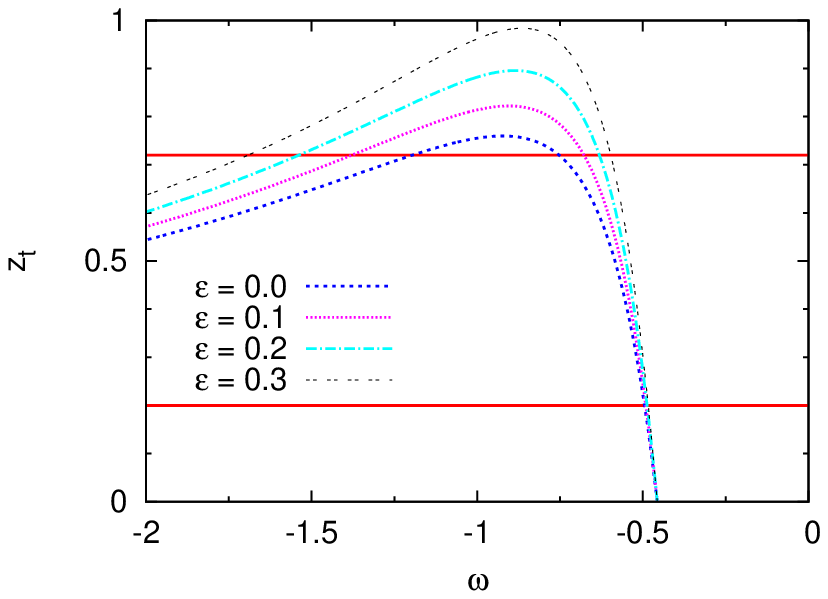,width=3.4truein,height=3.0truein}
\hskip 0.2in \epsfig{figure=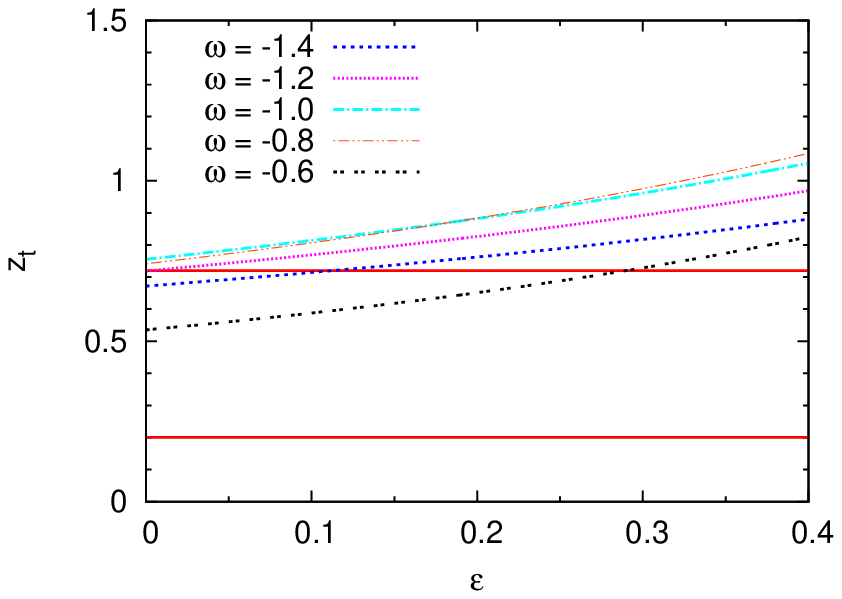,width=3.4truein,height=3.0truein}
\hskip 0.1in} \caption{The transition redshift $z_t$ in the scenario
of coupled quintessence. {\bf{a)}} $z_t$ as a function of $\omega$
for some  selected values of the coupling parameter $\epsilon$.
{\bf{b)}} The plane $(z_t - \epsilon)$ for some selected values of
$\omega$. In both diagrams, the horizontal lines represent the
$2\sigma$ limits on $z_t$ as given in Ref. \cite{rnew}.}
\label{fig:zt}
\end{figure*}

Now, by integrating Eq. (\ref{coupling}) it is straightforward to
show that the energy density of the dark energy component is given
by
\begin{equation}
\rho_x ={\rho}_{x0}{a}^{-3(1+\omega)}+
\frac{\epsilon\rho_{dm0}}{3|\omega|-\epsilon}{a}^{-3+\epsilon},
\end{equation}
where the integration constant ${\rho}_{x0}$ is the present-day
fraction of the dark energy density. Clearly, in the absence of a
coupling with the CDM component, i.e., $\epsilon=0$, the
conventional non-interacting quintessence scenario is fully
recovered. For $\omega=-1$ and $\epsilon\neq0$, we may identify
${\rho}_{x0}\equiv{\rho}_{v0}$ (the current value of the vacuum
contribution), and the above expression reduces to the vacuum
decaying scenario recently discussed in Refs.~\cite{wm,alclim05}.

Neglecting the radiation contribution, the Friedmann equation
(\ref{fried}) for this interacting dark matter-dark energy cosmology
can be rewritten as
\begin{equation}
\label{friedmann} {\cal{H}}= \left(\Omega_b{a}^{-3} +
\frac{3|\omega|\Omega_{dm}}{3|\omega| - \epsilon}{a}^{\epsilon - 3}
+ \tilde{\Omega}_{x}{a}^{-3(1 + \omega)}\right)^{1/2},
\end{equation}
where ${\cal{H}}= H(z)/H_0$, $\Omega_b$ and $\Omega_{dm}$ are,
respectively, the normalized Hubble parameter, and the baryons and
CDM present-day density parameters (for these quantities, we have
dropped the subscript `0' for convenience). The parameter
$\tilde{\Omega}_{x}$ is defined, in terms of the density parameter
of the  dark energy component $\Omega_{x}$, as
\begin{equation}
\tilde{\Omega}_{x} = \Omega_{x} -
\frac{\epsilon\Omega_{dm}}{3|\omega|-\epsilon}\;,
\end{equation}
and, therefore
\begin{equation}
\label{def_wx} \tilde{\Omega}_{x} = 1 - \Omega_{b} -
\frac{3|\omega|\Omega_{dm}} {3|\omega|-\epsilon}\;.
\end{equation}

The above expression clearly shows that the conventional
(non-interacting) quintessence scenario is considerably modified due
to the  dark energy decay into CDM particles. It is also worth
noticing the importance of the baryonic contribution to this sort of
scenario. Going back to high redshifts we see that the presence of
an explicit baryon term -- redshifting as $(1 + z)^3$ -- is well
justified since the decaying  dark enegy component slows down the
variation of CDM density. Although being subdominant at the present
stage of cosmic evolution, the baryonic content will be dominant (in
comparison to CDM) at very high redshifts. Actually, it becomes
subdominant just before nucleosynthesis ($z \simeq 10^{10}$ for
$\epsilon \sim 0.1$), so that the CDM component drives the evolution
after the radiation phase. However, even at late times, baryons have
several dynamical effects. In particular, they alter considerably
the transition redshift, i.e., the redshift where the current
accelerating regime begins \cite{alclim05}.

To quantify this latter effect, let us consider the transition
redshift, $z_t$, at which the Universe switches from deceleration to
acceleration or, equivalently, the redshift at which the
deceleration parameter vanishes. From Eq. (\ref{fried}), it is
straightforward to show that the deceleration parameter, defined as
$q(t) = -a\ddot{a}/\dot{a}^2$, now takes the following form
\begin{equation}
q(z)=
\frac{1}{2}\frac{1+\frac{3|\omega|+3\omega\epsilon}{3|\omega|-\epsilon}\frac{\Omega_{dm}}{\Omega_{b}}(1
+ z)^{-\epsilon} +
(1+3\omega)\frac{\tilde{\Omega}_{x}}{\Omega_{b}}(1 +
z)^{3\omega}}{1+\frac{3|\omega|}{3|\omega|-\epsilon}\frac{\Omega_{dm}}{\Omega_{b}}(1
+ z)^{-\epsilon} + \frac{\tilde{\Omega}_{x}}{\Omega_{b}}(1 +
z)^{3\omega}}\;.
\end{equation}
Some interesting features of the above expression must be explored.
First, if $\tilde{\Omega}_{x}=\epsilon=0$ one finds $q=1/2$, as
expected for a flat matter-dominated model
($\Omega_{b}+\Omega_{dm}=1$). Note that the expression of
$\tilde{\Omega}_{x}$ has been defined by Eq. (\ref{def_wx}). In
comparison with the conventional non-interacting quintessence
scenario, the coupling term ($\epsilon$) modifies considerably the
transition deceleration/acceleration. In principle, since the CDM
density scales as $a^{-3+\epsilon}$, while the baryon density scales
with $a^{-3}$, the latter becomes dynamically more important in
comparison with the non-decaying scenario. The overall baryon effect
is to delay the transition epoch relative to previous cases
(including the standard $\Lambda$CDM model), which seems to be in
better agreement with some recent SNe results indicating $z_{t} =
0.46 \pm 0.13$ at 1$\sigma$ \cite{rnew}. Besides this effect, we
also have in our model the effect of the dark energy EOS $\omega$,
which introduces a new parameter relative to the decaying vacuum
model \cite{alclim05}.
To better visualize the effect of the dark energy EOS, as well as
the effect of the dark energy-dark matter interaction parameter
$\epsilon$, we plot in Figure \ref{fig:qzw} the deceleration
parameter as a function of redshift $z$ for some selected values of
the EoS ($w = -1.1$ and $-0.8$) and decaying ($\epsilon = 0.1$)
parameters.

Figures (1a) and (1b) show that the effect of a positive $\epsilon$
parameter, as required by thermodynamical arguments \cite{alclim05},
is to decrease the value of the deceleration parameter, then
increasing the value of the transition redshift. The net effect of
the dark energy EOS, however, is not monotonic as the effect of the
interaction. We have in general, for values of the dark energy EOS
close to the vacuum value ($\omega=-1$), that the present value of
the deceleration parameter increases for increasing $\omega$,
although this behavior may change for values of $\omega$  much
different from the standard value, or at high redshifts, as seen in
Figure (1c).

Figures (2a) and (2b) show the direct effect of the $\epsilon$ and
$\omega$ parameters on the transition redshift ($z_t$). In agreement
with Figure (1), we see from Panel (2a) that the effect of the
interacting parameter is in general to increase $z_t$. We can also
see that the transition redshift has a maximum for values of the
$\omega$ parameter a little larger than $-1$. Similar result can
also be taken from Panel (2b), although $z_t$ has a more complicated
dependence on $\omega$. We also show in both panels the $2\sigma$
limits given by Riess {\it et al.} (2004) \cite{rnew}. It is clear
that values of $\omega$ larger than $-1$ as well as smaller than
$-1$ are favoured by these limits, whereas larger values of the
$\epsilon$ parameter are disfavoured by these limits when
$\omega\approx-1$. This latter result is in full agreement with the
results of our statistical analyses discussed in the next section.

\section{Analysis and discussion}

The description of the model discussed in the previous Section
clearly shows that it comprises a multitude of cosmological
solutions. In a model with such a wealth of different possibilities
constraints on the parameter space arising from current
observational data are likely to rule out many of the possible
scenarios (combinations of $\epsilon$, $w$ and $\Omega_{dm}$) for
the evolution of the Universe. In this Section we investigate such
observational constraints by placing cosmological bounds on the
parametric spaces $\epsilon-w$ and $\epsilon-\Omega_{dm}$ from
statistical analyses involving a large set of cosmological
observations. To this end we use the most recent distance
measurements to SNe Ia \cite{davis} and the current estimates of the
baryon acoustic oscillations found in the SDSS data \cite{bao}, as
well as, the shift parameter from WMAP observations \cite{Sperg07}.
In our analysis we fix $\Omega_b=0.0416$ also from WMAP results, a
value in good agreement with the constraints derived from primordial
nucleosynthesis \cite{nucleo}. Now, concerning the Hubble parameter, it
should be  recalled that the estimates of $H_0$ through different
methods fall on the range 62-74 km/s/Mpc with an uncertainty of
about $10\%$ \cite{H0}.  In what follows, we consider the Hubble
Space Telescope ($HST$) Key Project final result, i.e., $h =
0.71\pm0.08$ \cite{freedman}, as a Gaussian prior on the Hubble
parameter.

\subsection{SNe Ia}

The predicted distance modulus for a supernova at redshift $z$,
given a set of parameters $\mathbf{s}$, is
\begin{equation} \label{mag}
\mu_p(z|\mathbf{s}) = m - M = 5\mbox{log} d_L + 25,
\end{equation}
where $m$ and $M$ are, respectively, the apparent and absolute
magnitudes, the complete set of parameters is $\mathbf{s} \equiv
(H_o, \Omega_{dm}, \epsilon, w)$ and $d_L$ stands for the luminosity
distance (in units of megaparsecs).

\begin{figure*}
\centerline{\epsfig{file=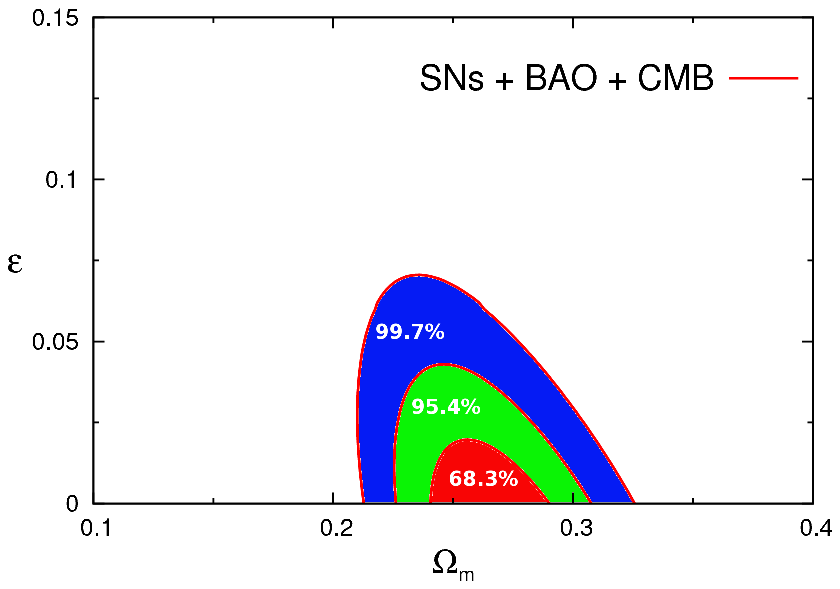,width=8.5cm,height=6.5cm}
\epsfig{file=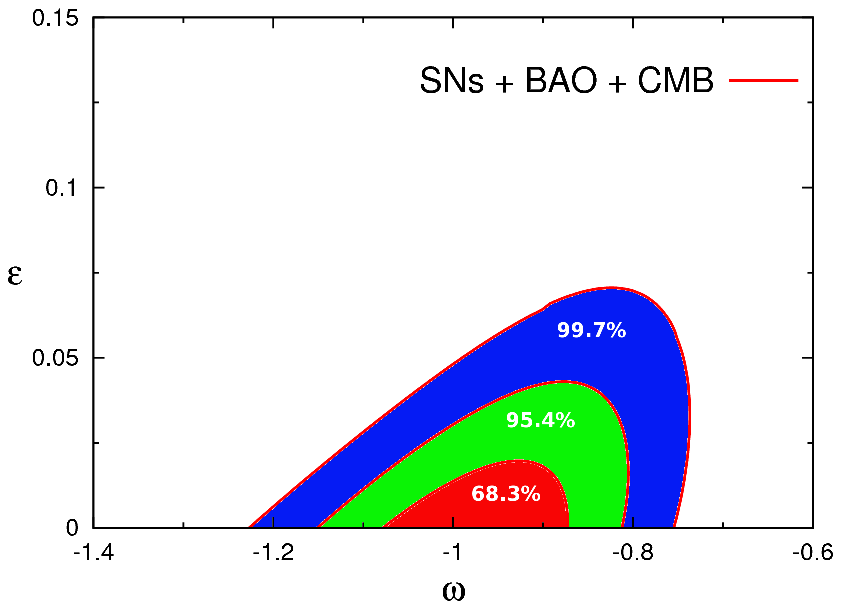,width=8.5cm,height=6.5cm}\hskip 0.1in}
\caption{The results of our statistical analyses. {\bf a)}
Confidence contours at 68.3\%, 95.4\% and 99.7\% in the plane
$\Omega_m$ - $\epsilon$ from a joint analysis involving SNe Ia + BAO
+ CMB shift parameter + $H_0$. As discussed in the text, this
analysis constrains $\epsilon$ to values very close to zero ($\simeq
0.09$ at 3$\sigma$). {\bf b)} Same as Panel {\bf a} for the plane
$\omega$ - $\epsilon$.}
 \label{fig:projhw}
\end{figure*}

We estimate the best fit to the set of parameters $\mathbf{s}$ by
using a $\chi^{2}$ statistics, with
\begin{equation}\label{xhi2def}
\chi^{2} = \sum_{i=1}^{N}{\frac{\left[\mu_p^{i}(z|\mathbf{s}) -
\mu_o^{i}(z|\mathbf{s})\right]^{2}}{\sigma_i^{2}}},
\end{equation}
where $\mu_p^{i}(z|\mathbf{s})$ is given by Eq. (\ref{mag}),
$\mu_o^{i}(z|\mathbf{s})$ is the extinction corrected distance
modulus for a given SNe Ia at $z_i$, and $\sigma_i$ is the
uncertainty in the individual distance moduli.

In our analysis, we use a combined sample with $N = 192$ SNs also
used by Davis {\it et al.} (2007) \cite{davis}. This sample consists
of the best quality light-curves SNs of Wood-Vasey {\it et al.}
(2007) \cite{wv07}, which are 60 ESSENCE supernovae \cite{wv07}, 57
SNLS supernovae \cite{Astier06}, and 45 nearby supernovae. We also
include, as in \cite{davis}, 30 new released SNe Ia, classified as
``gold'' supernovae by Riess {\it et al.} (2007) \cite{Ries07}.

\subsection{BAO}

The Baryon Acoustic Oscillations (BAO) given by the acoustic
oscillations of baryons in the primordial plasma, leave a signature
on the correlation function of galaxies as observed by Eisenstein
{\it et al.} (2005) \cite{bao}. This signature furnishes a standard
rule which can be used to constrain the following quantity
\cite{bao}:
\begin{eqnarray}
{\cal{A}} \equiv \frac{\Omega_m^{1/2}}{
{{\cal{H}}(z_{\rm{*}})}^{1/3}}\left[\frac{1}{z_{\rm{*}}}
\Gamma(z_*)\right]^{2/3}  = 0.469 \pm 0.017, 
\label{A}
\end{eqnarray}
where ${\cal{H}}$ is given by Eq. (\ref{friedmann}), $z_*=0.35$ is a
typical redshift of the SDSS sample, and $\Gamma(z_*)$ is the
dimensionless comoving distance to the redshift $z_*$. As has been
shown in Ref. \cite{baode}, this quantity can be used for models
which do not have a large contribution of dark energy at early
times.

\subsection{CMB shift parameter}

A useful quantity to characterize the position of the CMB power
spectrum first peak is the shift parameter, which is given, for a
flat Universe, by \cite{efstathiou}:
\begin{equation}
 {\cal R}=\sqrt{\Omega_m}\int_0^{z_r}\frac{dz}{{\cal H}(z)} = 1.71 \pm 0.03\;,
\end{equation}
where $z_r = 1089$ is the recombination redshift and the value for
${\cal R}$ above is calculated from the MCMC of the WMAP 3-yr in the
standard flat $\Lambda$CDM model \cite{ElgMult}.

As mentioned above, we also include a Gaussian prior on $h$, as
given by the final results of HST Key Project \cite{freedman}. Thus,
in our statistical analysis we minimize the following quantity:
\begin{eqnarray}
\label{chi22}
\chi^2 &=& \sum_{i=1}^{192}\left(\frac{\mu_{obs,i}-\mu_{th,i}}{\sigma_{\mu,i}}\right)^2 + \left(\frac{{\cal A}-0.469}{0.017}\right)^2 \nonumber \\
&+& \left(\frac{{\cal R}-1.71}{0.03}\right)^2 +
\left(\frac{h-0.72}{0.08}\right)^2\;. \label{chi2}
\end{eqnarray}

\subsection{Results}

In Figure (3) we show the main results of our statistical analyses.
As usual, the total likelihood is written as ${\cal L}\propto
e^{-\chi^2/2}$, where $\chi^2$ is given by Eq. (\ref{chi22}) . By
marginalizing ${\cal L}$ over the EoS parameter $\omega$, we can
quantify how much the plane ($\epsilon$-$\Omega_m$) can be
constrained by the data. The contour levels for this analysis are
shown on Figure (\ref{fig:projhw}a). At 68.3\%, 95.4\% and 99.7\%
c.l., we have found, respectively,
$$
\Omega_m=0.269^{+0.028 +0.047 +0.066}_{-0.026 -0.042 -0.058}
$$
and
$$
\epsilon=0.000^{+0.027 +0.057 +0.088}_{-0.000 -0.000 -0.000}\;,
$$
with the relative $\chi^2/\nu\simeq1.03$, where $\nu$ is the number
of degrees of freedom. These results are much more constraining than
those obtained in Ref. \cite{alclim05} for the case $w = -1$, as we
have used more recent CMB and SNe Ia data. While in the above
reference the bounds on the interacting parameter were
$\epsilon=0.06\pm0.10$ at $95.4\%$ c.l., we have found, at the same
level, $\epsilon=0.000^{+0.057}_{-0.000}$, which clearly constrains
this parameter to values very close to the standard non-interacting
case ($\epsilon=0$).

In Figure (\ref{fig:projhw}b) we show the plane ($\omega-\epsilon$)
when the total likelihood is marginalized over the density parameter
$\Omega_m$. For this analysis, we have found
$$
\omega=-1.006^{+0.117+0.188+0.258}_{-0.119-0.205-0.296}\;,
$$
whereas the bounds for $\epsilon$ are very similar to those found in
the previous analysis (Fig. 3a). Clearly, the standard $\Lambda$CDM
is preferred by this analysis, although much space is left for an
EoS distinct from $-1$. The so-called phantom models ($\omega<-1$)
are slightly more favoured by this analysis than quintessence
($\omega>-1$) scenarios.

\section{Final remarks}

The current standard cosmological model, i.e., a flat, accelerating
Universe composed of  $\simeq 1/3$ of matter (baryonic + dark) and
$\simeq 2/3$ of a dark energy component in the form of the vacuum
energy density ($\Lambda$), is fully consistent with a variety of
observational data. Even so, given the complexity of the involved
phenomena, it is clear that in order to obtain a deeper insight into
the nature of the dark energy and dark matter, it is worth consider,
both from the observational and theoretical viewpoint, more complex
scenarios as, for instance, models with interaction between these
two components.

In this paper we have discussed some cosmological consequences of an
alternative mechanism of cosmic acceleration based on a general
class of coupled quintessence scenarios whose interaction term is
deduced from the effect of the dark energy on the CDM expansion
rate. The resulting expressions for the model, parameterized by a
small positive parameter ($\epsilon$), are very general and have
many of the previous phenomenological approaches as a particular
case. In particular, the coupled quintessense models proposed here
may be thought as a natural extension of the decay vacuum scenarios
discussed a couple of years ago \cite{wm,alclim05}.

By combining the most recent SNe Ia, BAO, CMB shift parameter data
and the {\emph{HST}} results on $H_0$ we have shown that strong
constraints can be placed on this kind of scenario. We have shown
that the free parameters of the model are constrained to assume
values very close to the standard $\Lambda$CDM values, i.e., $\omega
\simeq -1$ and $\epsilon \simeq 0$, although space is still left for
an EoS distinct from $-1$ and the interacting parameter slightly
different from zero. It is worth emphasizing that in our analysis
the EoS $w$ and interacting $\epsilon$ parameters  have been set as
constants. In a more realistic case, however, such parameters must
vary with redshift. The theoretical and observational consequences
of this more realistic interacting $w(z)$ scenario, as well as a
full comparison with the case discussed in the present analysis,
will appear in a forthcoming communication.

\begin{acknowledgements}

The authors are grateful to F. E. M. Costa and J. V. Cunha  for
helpful discussions. JFJ and RCS are supported by CNPq. JSA is
supported by CNPq under Grants 304569/2007-0 and 485662/2006-0, and
JASL is partially supported by CNPq  and FAPESP (04/13668-0).

\end{acknowledgements}


\begin{thebibliography}{30}
\bibitem{Brons1} Historically, the possibility of
a time varying  $\Lambda(t)$ was first advanced in the paper
of Bronstein\cite{Brons2}. A summary of the evolving ideas can be seen in the review papers by Peebles and Ratra\cite{review}, and Lima\cite{review}.
\end{thebibliography}
\end{document}